\documentclass{eas}
\usepackage{graphicx}
\begin{document}

\title{Submm/FIR Astronomy in Antarctica \\ Potential for a large telescope facility} \thanks{The participants of the ARENA workshop on Submillimetre Far-InfraRed Astronomy from Antarctica that was held at CEA Saclay, France in 2007, are greatly acknowledged, especially the speakers, the session chairmen and the SOC members:  
Ph. Andr\'e (CEA Saclay), W. Ansorge (RAMS-CON), J. Braine (L3AB Bordeaux),  P. Calisse (Cardiff University), P. Cernicharo (DAMIR Madrid), C. De Breuck (ESO),  N. Epchtein (ARENA coordinator), E. Fossat (LUAN Nice), Y. Fr\'enot (IPEV), L. Giacomel (EIE), P. Godon (IPEV),  F. Helmich (SRON), F. Israel (Leiden), P. Lapeyre (Thales Alenia Space), R. Neri (IRAM), H. Olofsson (Onsala Space Observatory), P. Persi (INAF Rome), M. Piat (APC Paris), L. Rodriguez (CEA Saclay),  M. Sarazin (ESO),  R. Siebenmorgen (ESO), T. Stark  (CfA Harvard), J. Storey (UNSW Sydney), J.-P. Swings (Li\`ege University), M. Vaccari (Padova University),  C. Walter (CEA Saclay), M. Wiedner (Universitaet zu Koeln),  H. Zinnecker (AIP Potsdam)
  - http://irfu.cea.fr/Sap/Antarctica}

\author{V. Minier}\address{CEA Saclay/DSM/IRFU/Service d'Astrophysique, France. email: vincent.minier@cea.fr}
\author{L. Olmi}\address{Osservatorio di Arcetri, INAF, Italy and University of Puerto Rico}
\author{P.-O. Lagage}\sameaddress{1}
\author{L. Spinoglio}\address{IFSI, INAF Roma, Italy}
\author{G.A. Durand}\sameaddress{1}
\author{E. Daddi}\sameaddress{1}
\author{D. Galilei}\address{IASF, INAF Bologna, Italy}
\author{H. Gall\'ee}\address{LGGE, Grenoble, France}
\author{C. Kramer}\address{Universitaet zu Koeln, Germany}
\author{D. Marrone}\address{NRAO/University of Chicago, US}
\author{E. Pantin}\sameaddress{1}
\author{L. Sabbatini}\address{Dipartimento di Fisica, Roma 3, Italy}
\author{N. Schneider}\sameaddress{1}
\author{N. Tothill}\address{Dept of Physics, University of Exeter, Exeter, UK}
\author{L. Valenziano}\sameaddress{4}
\author{C. Veyssi\`ere}\address{CEA Saclay/DSM/IRFU/SIS, France}
\runningtitle{Minier et al.: Submm/FIR Astronomy in Antarctica \dots}
\begin{abstract}
Preliminary site testing datasets suggest that Dome C in Antarctica is one of the best sites on Earth for astronomical observations in the 200 to 500-$\mu$m regime, i.e. for far-infrared (FIR) and submillimetre (submm) astronomy. We present an overview of potential science cases that could be addressed with a large telescope facility at Dome C. This paper also includes a presentation of the current knowledge about the site characterics in terms of atmospheric transmission, stability, sky noise and polar constraints on telescopes. Current and future site testing campaigns are finally described.
\end{abstract}
\maketitle
\section{Submm/FIR astronomy context and Antarctica}
\subsection{A new generation of FIR/submm facilities}

Far-infrared/submillimetre (FIR/submm - 100 to 1000 $\mu$m) astronomy is the prime technique to study the Ôcold UniverseÕ and unveil the birth and early evolution of planets, stars and galaxies. It is a relatively new branch of astronomy at the frontier between IR and radio astronomy. FIR/submm continuum observations are particularly powerful to measure the luminosities, temperatures and masses of cold dust emitting objects because dust enshrouded star-forming regions emit the bulk of their energy between 60 and 500 $\mu$m. The submm/FIR range of the spectrum or THz regime is also rich in several  lines that are  the only means to study the kinematical structure of the interstellar medium (ISM) of galaxies. They allow to probe different physical regimes, i.e. regions of widely different densities and temperatures, depending on their excitation levels and critical abundances. Observations at these wavelengths with a large telescope will primarily lead to breakthroughs in the study of star formation at all scales and understand its cosmic history back to the early Universe as well as in the understanding of galaxy evolution. Asteroids, Debris disks, Planet formation, Dust origin in evolved stars, Interstellar dust and Polarisation of dust in the Universe are also potential science drivers for FIR/submm astronomy. An overview of FIR/submm astronomical science at Dome C was presented at the Saclay ARENA workshop in June 2007 and is reviewed in the present paper. 

What is the context today of submm astronomy ? Two major submm facilities will become available in the coming years: the Herschel Space Observatory, a FIR/submm (60-500 $\mu$m) telescope in Space and ALMA, a ground-based mm-wave (350 $\mu$m-7 mm) interferometer on the Chajnantor plateau in the northern Atacama desert. Both facilities will have their specific niches. Herschel will have the ability to carry out large area imaging surveys of both the distant Universe (Franceschini 2001) and the nearby interstellar medium in our own Galaxy (Andr\'e \& Saraceno 2005). ALMA will make possible ultra deep searches for primordial galaxies (Blain 2001), as well as detailed kinematical investigations of individual protostars (Evans 2001). However, both Herschel and ALMA will have their own limitations. The Herschel telescope (3.5 m) will suffer from its only moderate angular resolution, implying a fairly high extragalactic confusion limit (Oliver 2001) and preventing the study of individual protostars in all but the nearest star-forming clusters of our Galaxy. ALMA will suffer from a small field of view ($10''$) and limited observable conditions in the FIR/submm, making extensive wide-field mapping impossible given the amount of time necessary to cover large star-forming complexes and fields of primordial galaxies. Beside these two major facilities, there is a constellation of submm/FIR telescope projects in operation or in study, aboard balloons (e.g. BLAST, OLIMPO, PILOTE etc), aboard an airplane (SOFIA), aboard satellites (e.g. Akari; SPICA) and on high-altitude plateaux and mountains (e.g. APEX; CCAT, a 25-m telescope project). For instance, the APEX telescope at Chajnantor could allow 450-$\mu$m observations as recently demonstrated in 2007 by the CEA ArT\'eMiS project. BLAST , the Balloon-borne Large Aperture Submillimeter Telescope, observes simultaneously in 3 broadband filters at 250, 350 and 500 $\mu$m with detector arrays, from an altitude of 40 km. Two science flights have been performed: a 4-day flight from northern Sweden in 2005 and an 11-day flight in Antarctica in 2006 (e.g. Chapin et al. 2008).

Why evaluating potential science with a FIR/submm telescope at Dome C ? Beside Herschel and ALMA, there is thus a clear need for a large ($>10$ m) single-dish telescope (1) operating at 200-450 $\mu$m and providing (2) better angular resolution than Herschel and (3) wider-field mapping capabilities than ALMA, making large-scale mapping with a relatively good angular resolution ($\sim1''$) possible and well matched with thermal infrared space telescope (e.g. Spitzer).  New sites are therefore intensively tested because the 200-350-450-$\mu$m windows at Chajnantor open less than $30\%$ of wintertime at an observable level, probably less than $10\%$ at 200 $\mu$m. The stability of the atmosphere is an equally important parameter when comparing the sites and Dome C may stand out as being far more stable than Chilean sites (Minier et al. 2007 and reference therein). Equipped with FIR/submm imagers and spectrometers, a European telescope at Dome C in Antarctica might be able to operate in all atmospheric windows between 200 $\mu$m and 1 mm, and very regularly at 450 $\mu$m in wintertime.  As a demonstration Antarctica is indeed a very good site for submm/mm astronomy, US astronomers have built a submm telescope of 10 m (the SPT) at the South Pole (Ruhl et al. 2004). On a higher and more stable site, Dome C could become the European observatory in Antarctica.

\subsection{Dome C in Antarctica: a potential site for FIR/submm astronomy}

Dome C is the location of the French-Italian Concordia station that is connected to Dumont-D'Urville on the coast by either light planes or ground motorised raids for transporting heavy material. The Station building can now host $\sim15$ people during winter and therefore allows experiments all year long \footnote{http://www.concordiastation.org/}. Site testing and qualification of the site for optical and near-infrared astronomy have been conducted for many years by French, Italian and Australian teams. However, little efforts have been produced so far to evaluate the quality of the atmosphere, the climate constraints and the specificity of Dome C for a potential submm/FIR telescope. Calisse et al. (2004) undertook the measurement of the atmosphere opacity at 350 $\mu$m during a summer period and found it comparable to that at South Pole. No direct assessment of the wintertime atmosphere transmission has ever been performed. 

\begin{figure}
\centering
\includegraphics[scale=0.6]{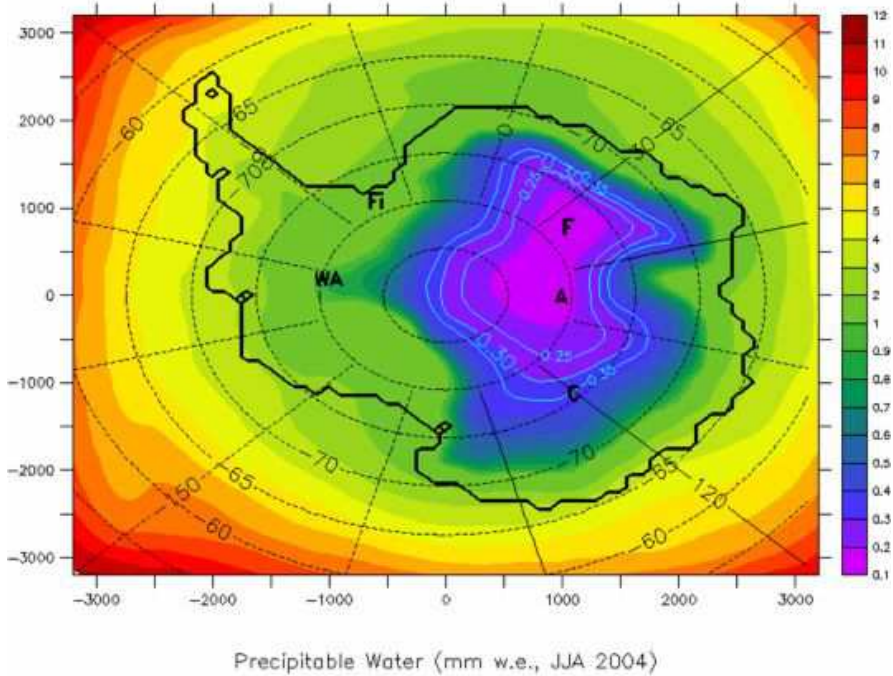}
\caption{Precipitable Water Vapour content (PWV) distribution over Antarctica as simulated by the regional climate model MAR (june-august 2004 average). Mean PWV at Dome C is around 0.35 mm while at Dome A it is around 0.2 mm. From Gall\'ee, 2008, Monthly Weather Review, in prep.}
\end{figure}

A major obstacle to carry out submm observations from ground is the water vapour in the atmosphere. Astronomical observations in the FIR/submm spectral bands (e.g. 200, 350, 450 $\mu$m) can only be achieved from cold, dry and stable sites with ground-based telescopes or from space to overcome the atmosphere opacity and instability that are mainly due to water vapour absorption and fluctuations in the low atmosphere. Chile currently offers the driest accessible (all-year long) sites on Earth, where the precipitable water vapour (PWV) content is often less than 1 mm. The Chajnantor plateau (5100 m) hosts the ESO facilities for submm astronomy: the APEX 12-m telescope and the coming ALMA interferometer. However, FIR/submm observations at 200, 350, 450 $\mu$m are only possible when PWV drops below 0.5 mm, which occurs less than $30\%$ of wintertime at the ALMA site. In addition, observing conditions in Chilean sites can progressively and frequently be degraded by climate phenomena like the Bolivian winter or El Ni\~no that brings weather instabilities. Possibly global warming has a more severe impact on these sites than on Antarctica. Other potential Chilean sites at higher altitudes (e.g. Cerro Chajnantor, Sairecabur) presently undergo site testing. 

The plateau of Antarctica might possibly be a very privileged and alternative area for achieving astronomical observations in the submm/FIR range. The geographical and climate conditions are extreme, which favour low PWV in the atmosphere: the low sun cover, the isolation by the circumpolar stream of the Antarctic ocean and the high power of reflection of ice make Antarctica the coldest continent on Earth. It is also important to emphasize that snow precipitation is very low on the Antarctic plateau. The low pressure fronts do not penetrate into the inner plateau and remain located on the coast lines. In fact the inner part of the continent is a true desert: on an area of 5 millions of km$^2$, snow precipitation is about 5 cm, and often less than 2 cm on the highest Domes. As a consequence, the PWV at Dome C is very low and expected lower than at Chajnantor in average (Fig. 1). However, a proper comparison between Dome C and other sites should be based on the comparison of the transmission, the linewidth of the atmospheric windows and their stability, and the level of skynoise. The transmission and the linewidth of the windows mainly depend on the PWV, temperature and pressure, while the sky noise depends on the water vapour cell fluctuations in the low atmosphere (Fig. 2). 

Preliminary site testing in summertime and modelling predict that the French-Italian Concordia base at Dome C in Antarctica is a potentially remarkable site on Earth for FIR/submm astronomy. Measurements of the humidity at different altitudes with radiosounding techniques (Fig. 3) and derivation of the PWV (Fig. 4) over a statistically valid time tend to demonstrate that Dome C is a slightly better site than the best Chajnantor site (i.e. cerros or mountains at $>5500$ m) in terms of PWV percentiles. However, when taking into account the temperature and pressure effects, it is not clear that Dome C is a better site in terms of overall transmission (amplitude, linewidth and time percentiles; see inset in Fig. 2). In addition, the PWV does not fall below 0.2 mm very often, which is critical for opening the 200-$\mu$m windows. Nonetheless the atmospheric transmission and stability are probably always ideal for observations at 350 and 450 $\mu$m. For non-exclusive use of telescope time for these wavelengths, the Chajnantor plateau is to date  the optimal site as submm observatories are already in place. For regular ($>100$ days) use of telescope time for the 350-450 $\mu$m windows and complementary observations at 200 $\mu$m, Dome C might become, however, a much more attractive site. 

A complete assessment of the 200-$\mu$m  opacity of the atmosphere is therefore crucial for concluding whether Dome C is better than any other known sites in the Chilean Andes and more generally on Earth. Skynoise will also be measured by CAMISTIC  in 2010 (Minier et al. 2007). Note that site testing experiments at Dome A will be undertaken (PLATO project) in 2008 with preHeat a specific experiment that is dedicated to assess the site for submm astronomy (Tothill et al. this volume). Another crucial assessment is the understanding of the effects of temperature gradients in the inversion temperature layer between 0 and $\sim30$ m and icing formation during polar nights on telescope hardware and operation (Fig. 5). Nighttime cooling of ice sheet by radiation to space creates a powerful temperature gradient and generates near surface level inversion winds where air, cooled by contact with the surface, flows down the gently sloped interior of the plateau (Swain \& Gall\'ee 2006). The studies of these constraints are realised under the GIVRE experiments (Fig. 5), which are currently in operation at Dome C.

\begin{figure}
\centering
\includegraphics[scale=0.8]{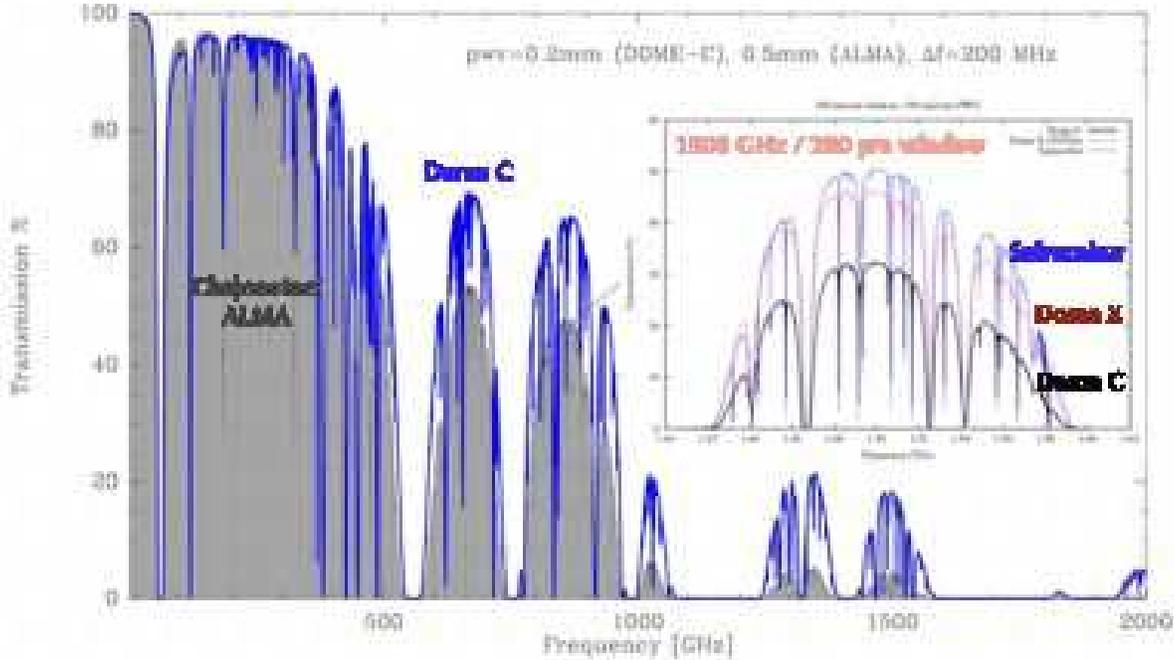}
\caption{Modelled transmission for Dome C and Chajnantor/ALMA site with the MOLIERE code. Transmissions were estimated using PWV=0.2 mm for Dome C (blue) and 0.5 mm for ALMA site (grey) that represent the expected first quartile PWV values. 200-$\mu$m (1500 GHz) windows open at Dome C. Note: MOLIERE (Microwave Observation LIne Estimation and REtrieval) is a versatile forward and inversion model for the millimetre and submillimetre wavelengths range, used in many aeronomy and some astronomy applications (Urban 2004). Plots produced by N. Schneider-Bontemps. Inset plot with {\it am} code shows the comparison of the 200-$\mu$m transmission for PWV=0.1 mm at Dome C, at Sairecabur (a 5525-m Chilean site) and Dome X, a fictive site with atmospheric profile as Dome C but at a pressure altitude of 5525 m. The Chilean site has the highest transmission. At equal altitudes and pressures, Chilean site vs. Dome X, Chilean site would win over Antarctica lower temperature conditions. This is because low temperatures bias H$_2$O partition function and strengthen THz absorption lines. Comparing Dome C with Sairecabur, higher atmospheric pressure at Dome C broadens very strong lines that bound THz windows. Overall: the transmission for PWV=0.2 mm at Dome C is comparable to the transmission for PWV=0.35 mm at 5500m in Chile. However, this comparison does not take into account the percentiles of PWV at both sites and need to be confirmed by observations. The models were done with the am modeling software: http://sma-www.cfa.harvard.edu/private/memos/152-03.pdf. Plots produced by D. Marrone}
\end{figure}

\begin{figure}
\centering
\includegraphics[scale=0.4, angle=90]{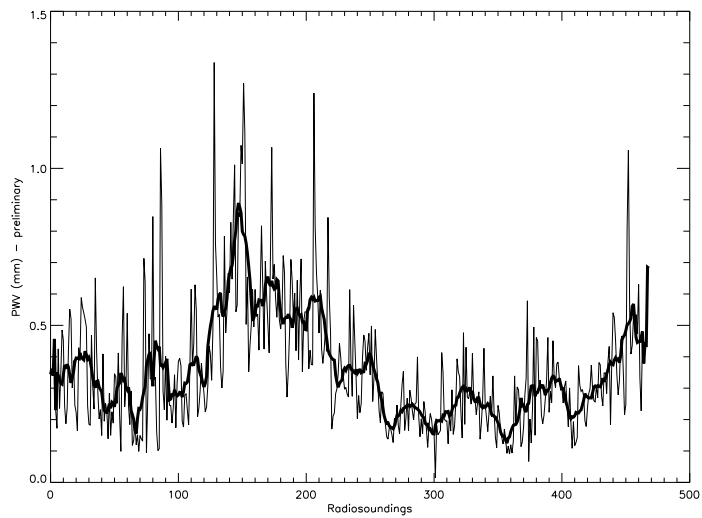}
\caption{PWV time series at Dome C from radiosounding data (preliminary
results, dataset Apr. 05 - Jan. 07). A smoothed profile (thick line) is
overplotted. A significant variability is present in the dataset, which
demands further study. Produced by L. Valenziano.}
\end{figure}

\begin{figure}
\centering
\includegraphics[scale=0.4, angle=90]{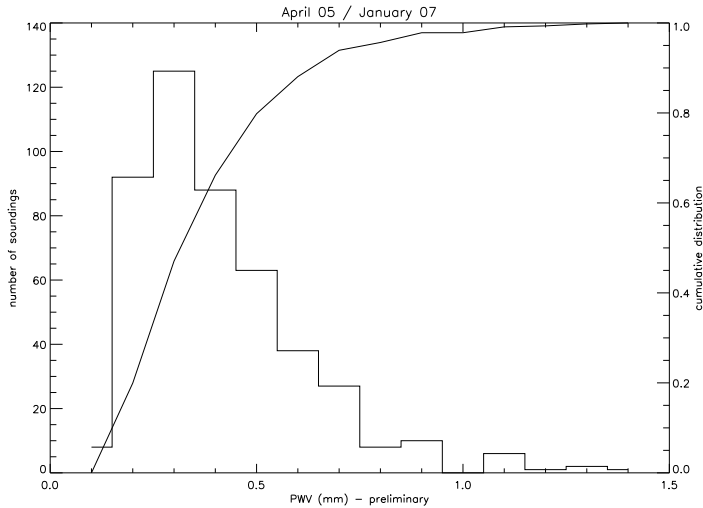}
\caption{ Precipitable Water Vapor distribution at Dome C from
radiosounding data (preliminary results, dataset Apr. 05 - Jan. 07),
corrected according to Tomasi et al. 2006. The cumulative distribution
is also shown. We report the 25th, 50th and 75th quartiles compared with
the best submillimetre sites around the Chajnantor plateau up to 5800 m (Giovannelli et al.
2001): 0.23 (0.27), 0.33 (0.49), 0.48 (0.92). Produced by L. Valenziano.}
\end{figure}

\begin{figure}
\centering
\includegraphics[scale=0.45]{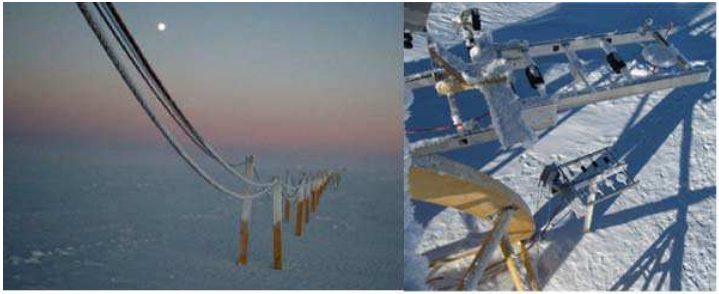}
\caption{Left: Formation of frost on sticks during the polar winter. Note that the base of the sticks are frostfree while most icing forms on the top. Right: GIVRE experiment by CEA Saclay at Dome C. GIVRE (=icing in French) measures the formation of frost on disks, cylinders made from different material. Icing always forms on each type of material. This is likely due to the general Dome C climate conditions: the combination of little wind and great heat loss  by radiation at ice level during the polar winter results in a large thermal inversion and strong icing on all hardware objects. GIVRE is CEA led experiment under G. Durand's coordination in partnership with LUAN, IPEV and Roma 3.
 }
\end{figure}

\section{Science drivers}
\subsection{Galaxy formation and evolution}

The Cosmic Infrared Background (CIB) was discovered with the COBE satellite a decade ago. It results from the emission in the infrared (IR) of distant galaxies, the so-called infrared galaxies. These galaxies appear to emit more intensively in the IR than the most nearby galaxies. Two physical mechanisms may explain the high level of IR emission that is measured in the CIB. First the distant galaxies may form stars in a much more productive ways, through starburst mechanisms. The UV lights of high mass stars heat up the dust grains in the interstellar medium that re-emit in the IR and submm. Consequently, the Spectral Energy Distribution (SED) of these IR and submm galaxies peaks in the submm/FIR due to star formation (Fig. 6). A second mechanism may be the effect of supermassive black holes in the active galactic nuclei (AGN) that emit in X-ray after swelling hot gas from nearby disrupted stars. Dust again absorbs the X-ray radiation and re-emit in the IR. The IR excess due to AGN manifests at 24 $\mu$m mainly (Fig. 6; Daddi et al. 2007). In both cases, the CIB provides astronomers with information on the galaxy formation and evolution, as well as on the history of cosmic star formation in the Universe. 

The CIB in the submm/FIR is still largely unexplored territory as compared to radio and optical/IR sky where all-sky maps of the background exist. Only small areas ($1^o\times1^o$) have been mapped up to date. Observations with Akari and Spitzer, and then with Herschel should modify the current knowledge of the CIB and high-angular resolution observations of the new fields will be needed. Two major tasks could be accomplished in FIR/submm in order to progress in the understanding of the CIB and galaxy formation and evolution: To cense star formation in the distant Universe, to account for most of stellar mass in galaxies; To cense obscured (by dust) AGN activity by determining the bolometric luminosity of the associated IR galaxy. Major questions are: How massive ellipticals were formed, through merging ? What is the stellar initial mass function (IMF) for galaxy formation at high-z ? At what epoch(s) did most of the star formation takes place? What is the link between AGN and galaxy activity ?

Observing in the submm/FIR at Dome C is an exciting opportunity if a telescope facility could resolve galaxies at $1<z<3$ with a mass down to  $\sim10^{10}$ M$_{\odot}$. It will allow a proper estimate of their bolometric luminosity (see SED inset on Fig. 6).  This implies to go down to mJy sensitivities at submm/FIR wavelengths. Such observations would be possible with a 25-m aperture telescope at 200 and 450 $\mu$m under PWV$<0.5$ mm  that would allow deep field searches, resolving most of the infrared galaxies and providing  a complete census of star formation and AGN up to a redshift of z=3. This will be a very important breakthrough in the study of galaxy evolution.

\begin{figure}
\centering
\includegraphics[scale=0.8]{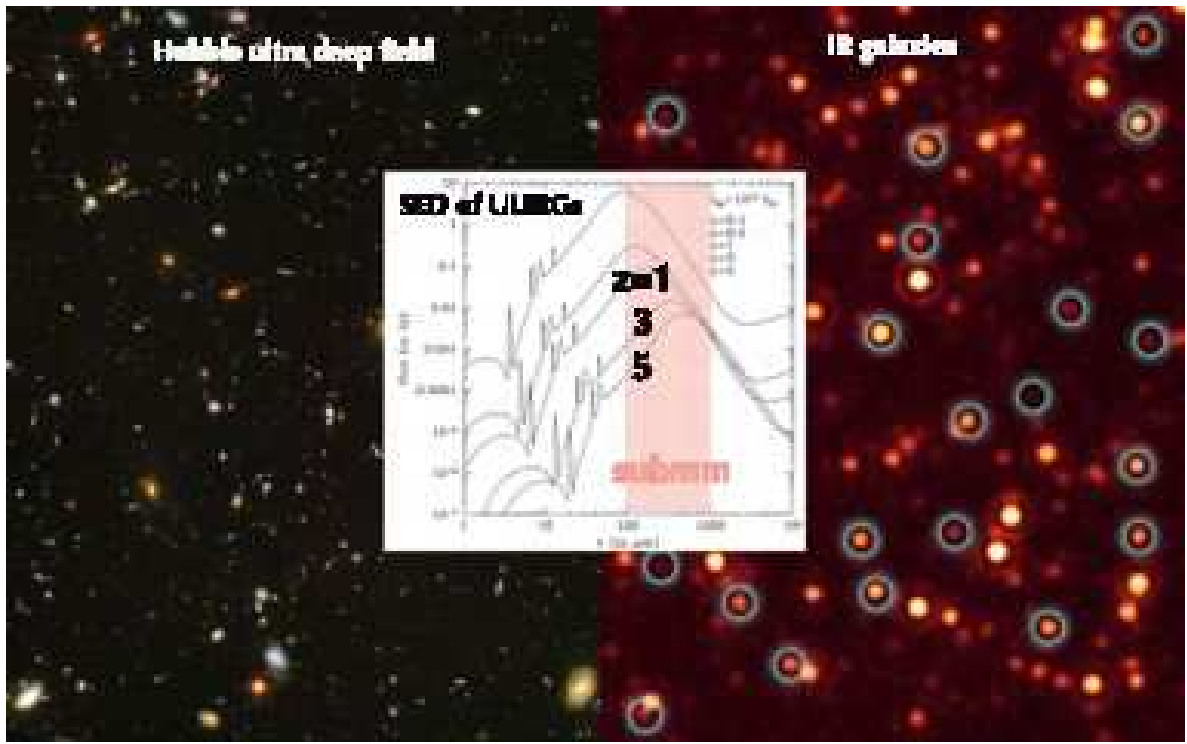}
\caption{Left: Hubble ultra-deep field (credit NASA). Middle: Spectral Energy Density of ultra-luminous galaxies for redshift z=0.1 to 5. Submm/FIR continuum observations measure SED peaks for z=3 to 5 (Guiderdoni et al. 1998). Right: A long-lost population of active supermassive black holes, or quasars, uncovered by the Spitzer and Chandra space telescopes. This IR image shows a fraction of these black holes, which are located deep in the AGN of distant, massive galaxies (circled in blue). This work is part of a multiwavelength program called the Great Observatories Origins Deep Survey, or Goods.  The Spitzer data show that hundreds of the galaxies between 9 and 11 billion light-years away shine with an unexpected excess of infrared light. X-ray data from Chandra of the same field reveal that the infrared-bright galaxies hide many black holes. This excess infrared light is probably being produced by the growing black holes (Daddi et al. 2007). }
\end{figure}

\subsection{Star formation up to the Magellanic Clouds and beyond}

The problem of the origin of the distribution of stellar masses at birth, or Ôinitial mass functionÕ (IMF), is one of the most important open issues in astrophysics today. It is not only central in local star formation research, but also crucial to understand whether the IMF is truly universal, including in starburst galaxies and at high redshift, or likely to depend on local physical conditions such as metallicity, pressure, or temperature. As prestellar cores and young (Class 0) protostars emit the bulk of their energy longward of 100 $\mu$m (Andr\'e et al. 2000), submm continuum mapping is a unique tool to address this problem. Recent ground-based dust continuum surveys of nearby, compact cluster-forming clouds ($\rho$-Ophiuchi, Orion B) at 850 $\mu$m and 1.2 mm suggest that the IMF of solar-type stars is largely determined by pre-collapse cloud fragmentation, prior to the protostellar accretion phase (i.e. the embryonic star phase). These results are, however, seriously limited by small-number statistics at both the low- and high-mass ends of the mass spectrum due to both sensitivity and angular resolution limitations of current telescopes. High-mass (M$_*>$8 M$_{\odot}$) stars are apparently able to form only in closely-packed stellar protoclusters that are mainly located in large, high-mass star-forming complexes within the Galactic plane at distances greater than 1 kpc (Minier et al. 2005). The lack of angular resolution around 200 $\mu$m up to now has essentially precluded the determination of accurate luminosities and dust temperatures for high-mass pre- and protostellar objects. With an angular resolution 3.5 times better than Herschel at $\sim$200-450 $\mu$m, a 12-m telescope can directly image individual protostars and prestellar condensations up to 2 kpc, providing access to the high-mass regime of the protostellar core mass and luminosity functions for the first time. A total of $\sim$700 prestellar precursors to OB stars are expected within 2 kpc of the Sun as opposed to only 20 high-mass prestellar condensations at d$<0.5$ kpc. Very southern star-forming regions (e.g. Chameleon, Carina) would be potential targets from Dome C in synergy with ALMA observations. The enhanced angular resolution of a ground-based antenna will also be well matched with that of space-borne mid-IR instruments.

The gain in angular resolution with a large ground-based telescope and the possibility to observe from 200 to 450 $\mu$m at Dome C could also foresee observations of star formation beyond the Milky Way. In the Magellanic Clouds (LMC and SMC), the angular resolution of a 12-m telescope at 200 $\mu$m would allow observation with a spatial resolution of $\sim1$ pc. Dome C has also a privileged location to observe the Magellanic Clouds (RA=05h Dec=-69$^{\mathrm{o}}$) and study the star formation rate in their gas-rich and metal-poor environment. 130 giant molecular clouds have been identified in the LMC and SMC with the NANTEN CO surveys. Further away, studies of the interstellar medium and star formation in starburst and interacting galaxies such as M82 or the Antennae would also be possible.

\begin{figure}
\centering
\includegraphics[scale=0.37]{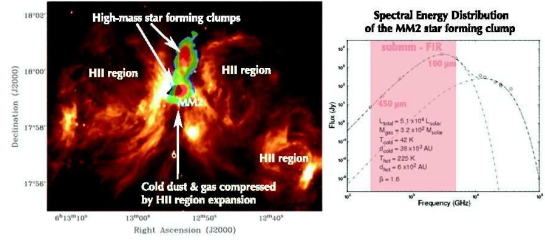}
\caption{Left: Spitzer IRAC image of thermal dust emission and PAH emission around HII regions. The RGB image overlaid on the Spitzer image is the cold dust emission from star-forming clumps as detected by the p-ArTeMiS bolometer array on APEX at 450 $\mu$m. Right: SED diagram of one of the star-forming clumps that is due to dust emission from 1 mm to 8 $\mu$m. The cold component is mainly constrained by the submm/FIR emission of dust that allows to estimate the luminosity and dust temperature, and therefore deduces the clump mass. Each clump is a progenitor of a star cluster or high-mass protostar (Minier et al. 2005).}
\end{figure}

\subsection{Molecular spectroscopy}

The submm/FIR range of the spectrum or THz regime is rich in
lines.  Observations
of spectral lines are the only means to study the kinematical
structure of the ISM. They allow to probe different physical regimes,
i.e. regions of widely different densities and temperatures, depending
on their excitation levels and critical abundances. Examples of spectral lines are given in Figure 8.

The [NII] 205-$\mu$m  fine structure line is one of the most important
cooling lines of the Milky Way as the COBE observations have shown (Fixsen et al. 1999). However, only very few ground-based observations of this
tracer of the warm ionized medium exist to date. The 205 $\mu$m [NII] line from the Carina Nebula was indeed detected with the AST/RO telescope at the South  Pole (Oberst et al. 2006). The [CII] line at 158 $\mu$m is also very important, but not observable from the ground. It could, however, be observed when redshifted into the 200 $\mu$m windows, from extragalactic sources. The total luminosity of [CII] from the Galaxy as a whole as measured by the COBE satellite is $5\times10^7$ L$_{\odot}$ and L[C II]$\sim$0.3\% of the FIR luminosity for star forming galaxies like the Milky Way. [CII]is the dominant coolant (also [CI] and [OI]) for much of the neutral ISM, including atomic clouds, the warm neutral medium, and PDRs. 

Other important cooling lines of the Milky Way ISM are the fine
structure lines of atomic carbon and the rotational lines of CO,
tracing dense molecular clouds and their surfaces. Depending on the
transition, the mid- and high-J CO lines trace the hot molecular gas
near sites of active star formation. The high angular resolutions
achieved with a 10m-class submm telescope will allow to study the inner,
hot parts of star-forming clumps and protostellar disks. Still higher
densities are sampled by the rotational transitions of HCN and HCO$^+$.

Far-infrared observations will allow to study the low-lying rotational
transitions of light hydrides HCl, SiH, CH, NH$_2$, NH$_3$ and their
deuterated species. These molecules represent a very important
component for the study of the innermost prestellar cores. In fact,
the densest parts of the pre-stellar nuclei are almost completely
deprived of molecules in the gaseous phase, especially CO, CS, but
even N$_2$H$^+$ and related isotopomers (e.g. Caselli et al. 1999). Since the H$_2$D$^+$ and D$_2$H$^+$ are probably the only
molecular species that can be observed on a scale of a few thousands
AU, they will play a fundamental role in the study of the physical
conditions and of the gas kinematics in those regions.  One of the main
goals of future observations of pre-stellar nuclei therefore consists
in determining the total abundance (ortho + para) of H$_2$D$^+$ in
order to characterize the present chemical models. This will be
possible through observations of the fundamental transitions of the
para-H$_2$D$^+$ around 1.4 THz that is not observable with Herschel and
of the D$_2$H$^+$ in the 200-$\mu$m atmospheric window. NH$_2$ and
NH$^+$ (296 $\mu$m) has not yet been detected in the ISM; they could
potentially probe collapsing pre-stellar cores.

Low bending modes of carbon chains (C8, C9, and C10 in THz) are
expected to be detected in circumstellar envelopes. Finally, for the
study of intermediate to high redshift galaxies, a variety of atomic
mid-IR lines from [NeII], [NeIII], [NeV], [SIII], [OIV], etc will appear at
different redshifts in the atmospheric windows which become open at Dome C.

\begin{figure}
\centering
\includegraphics[scale=0.45, angle=-90]{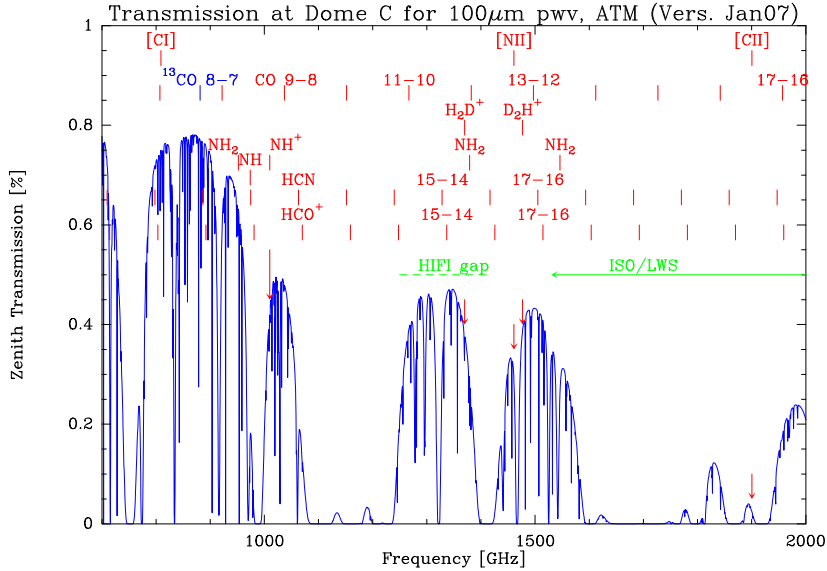}
\caption{Available spectral lines in the submm/FIR/THz regime in comparison with the transmission for PWV=0.1 mm at Dome C. Plot produced by C. Kramer.}
\end{figure}

\subsection{Other potential cases: planet formation; clouds in galaxies; SZ effect}

Submm/FIR astronomy with large aperture telescope would allow a deep study of the structure of protoplanetary and debris disks. Such studies would address key questions: What is the dust/gas ratio as a function of time? When does the dust settling/coagulation start? When does the gas phase disappear? What is the chemistry in these disks? At what ages of the systems did the planets form? Figure 9 illustrates the potential of observations at 200 $\mu$m that could constrain the outer disk radius, which is very difficult to determine in the mid-IR because of the lack of sensitivity to very cold dust.  Given the angular resolution needed ($\sim1''$), a very large aperture single dish ($>12$ m, $<4''$ at 200  $\mu$m) or a compact interferometer is required. 

Similarly, the study of individual clouds in nearby galaxies through [CI] lines and the 200/450 $\mu$m continuum that is ideal for cold dust at 6-8 K would need high angular resolution to be able to go down to the 15 pc spatial sizescale, i.e. a few arcsec at M31 and M33 distances. A telescope aperture of $\sim25$ m is therefore needed.

Cosmic Microwave Background photons interact with hot gas in clusters of distant galaxies. This interaction produces the Sunyaev ZelÕdovich (SZ) effect: low-frequency photons are kicked up to higher frequencies. The SZ energy distribution peaks at 2 mm and 750 $\mu$m (Fig. 10). The South Pole Telescope has been built to study SZ effect at 2 mm and studies of the "positive" part of the energy distribution might be possible at Dome C down to 450 $\mu$m with higher angular resolution.

\begin{figure}
\centering
\includegraphics[scale=0.4]{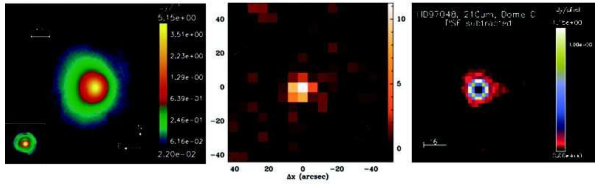}
\caption{Left: VLT/VISIR image of the HD97048 protoplanetary disk at  8.6 $\mu$m, showing its flared, extended surface (Lagage et al. 2006). The lower left inset is the VLT/VISIR PSF as a comparison. Middle: Observation of HD97048 with p-ArTeMiS/APEX at 450-$\mu$m with a pixel scale of $6''$. Imaged produced by Ph. Andr\'e. Right: the HD97048 emission at 200 $\mu$m has been simulated using a radiative transfer code, and using Dome C expected sensitivity. A scaled PSF observation has been removed, indicating that the disk image is resolved at 200 $\mu$m. This shows the potential to constrain the outer disk radius, which is very difficult to determine in the mid-infrared because of the lack of sensitivity to very cold dust. Image produced by E. Pantin.}
\end{figure}

\begin{figure}
\centering
\includegraphics[scale=0.4]{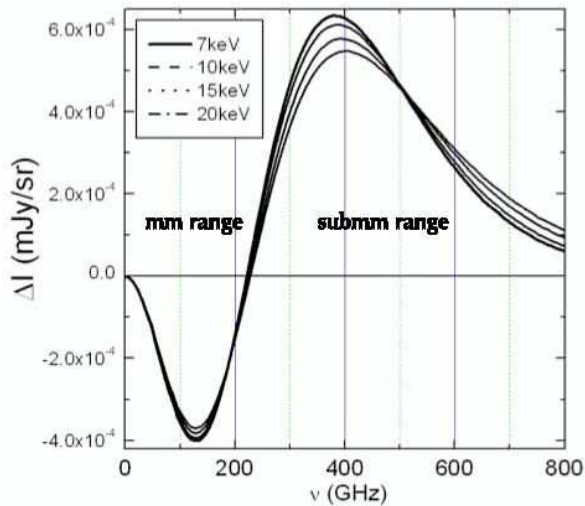}
\caption{SZ effect. Predicted flux distribution along the submm-FIR range of the spectrum. Extrema are at 2 mm and 750 $\mu$m. Courtesy of P. de Bernardis.}
\end{figure}

\section{Which telescope facility at Dome C ?}

\subsection{Future potential facilities}

Preliminary specifications on a potential telescope facility at Dome C are based on the above science driver requirements. In all science cases, a relatively high angular resolution ($\sim1''$) is required to address key questions. This is the most important gain with respect to space telescopes that are still limited in diameter. Observations at 200-450 $\mu$m are also fundamental for constraining key physical parameters such as the luminosity and temperature from dust emission SED. Finally the galaxy evolution and star formation cases require large areas ($\sim1^o\times1^o$) to be mapped. While ALMA can supply high angular resolution (e.g. for the disks) and will mainly study gas properties with spectral lines, ALMA cannot produce large area maps at 200-450  $\mu$m in a reasonable amount of time due to the limitation by the primary beam size ($\sim10''$) and atmosphere transmission. It cannot reach the THz (1000-1500 GHz) spectral line regime either. A large aperture telescope ($\sim25$ m) such as that proposed in the CCAT project for Cerro Chajnantor  would fulfill all the above requirements. A better site than Cerro Chajnantor is required to perform regular, ground-based observations at 200 $\mu$m. An alternative possibility would be an array of medium-size telescopes in a very compact baseline configuration that would simulate a 25-m dish sensitivity and angular resolution. Dome C might offer the best alternative on Earth to reach the 200-350-450-$\mu$m windows. However, the extreme difficulty to work at Dome C argues for simple, robust, and unmanned experiments, especially after February when winter-over staff cannot maintain instruments outside, and the temperature rapidly drops to -60 $^o$ C. Work on site must then be reduced to the installation and maintenance in the short summer period and operation in polar wintertime.

To date, there are different options for implementing a submm/FIR telescope facility at Dome C. First, an initial approach will exploit the available IR and mm telescopes (e.g. 80-cm IRAIT, 2.6-m COCHISE) on site and equip them with submm/FIR bolometer and heterodyne receivers. The goal would be to have a final assessment of the site characterization for these wavelengths. This is for instance the approach that is followed for the CAMISTIC 200-$\mu$m bolometer camera project (Minier et al. 2007). A more straighforward approach is the installation of a  winterized 12-m ALMA-type antenna that could both undertake unique world-class science and pathfinder experiments. It could also be a demonstrator for an interferometer. Such an interferometer with short baselines could be a THz heterodyne interferometer (no competition with ALMA) or a direct bolometric interferometer that could help in the development of technology for future Far-IR Interferometer space mission. The first approach will slowly prepare future large projects by learning about polar telescope engineering, unmanned control, telescope foundations, wind and ice formation on mirrors, active surfaces at $T<-70$~$^o$C, stability and transparency of submm/FIR windows, and optical designs. Alternative energy supply would also be needed such as photovoltaic panels (polar summer), 50-m high wind  turbines and H$_2$ fuel cells (polar winter) to power observatory facilities and the Concordia station. The second approach, the installation of a telescope for submm astronomy, will be based on the opportunity that ALMA-type antennas are under construction and the experience of the 10-m SPT deployment at South Pole.

\begin{figure}
\centering
\includegraphics[scale=0.35]{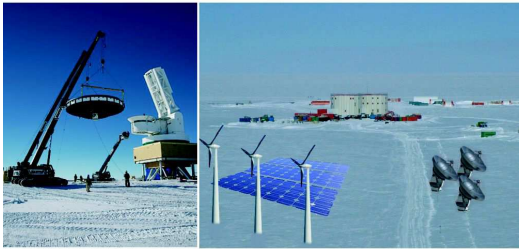}
\caption{Left: Deployment of the SPT at South Pole and required level of logistics. Right: View of a possible $3\times12$-m antenna interferometer at Dome C with renewable energy supply. The deployment of such telescopes at Dome C would require an important and costly upgrade of the current Concordia station.}
\end{figure}

\subsection{ALMA-type antenna as pathfinder}

The main technical rationale of installing a 12-m telescope at Dome C, or the Antarctic Submillimeter Observatory (ASO), is the availability of an existing optical and mechanical design for a 12-m submm antenna: the European ALMA prototype. This would dramatically cut the costs of developing a new antenna design ex novo. On the other hand, building an even larger antenna at Dome C, which would definitely benefit potential science results, without some pathfinder and well-developed logistical structures could prove very difficult and expensive. However, even the development of the 12m-telescope  would have to undergo a feasibility study for the winterization of the ALMA antenna that is required to operate under the extreme weather conditions of the Antarctic Plateau; this analysis should also address the pointing, tracking, and chopping requirements.  The pathfinder would represent the first step toward a larger facility instrument, either a bigger antenna, an interferometer, or even the modified ALMA antenna with a central surface to operate at mid-IR wavelengths.

The major advantages of ASO in terms of feasibility are that the basic antenna components will be available by 2011 and that operations could begin as soon as 2014. The limited increase in precision of the antenna, D/${\epsilon}$$\sim$10$^6$ (where D is the diameter of the primary reflector and ${\epsilon}$ is its surface RMS) compared to existing telescopes ensures that the required technology  upgrade would be kept to a minimum. This is very important since the technological efforts (and costs) to achieve the antenna specifications are not a simple linear function of D/${\epsilon}$. In fact, ASO-12m could be built at a small fraction of the CCAT-25m cost (estimated to be $\sim100$ MEuros). 


The potential use of the ALMA antenna as a submm as well as a mid-IR facility has been proposed by Olmi et al. (2007). The innermost surface of the ALMA prototype has a predicted RMS lower than the 12 micron expected across the whole 12 m diameter. However, it is also likely to be higher than that required to operate across the mid-IR window. Therefore,  a technique would have to be eventually devised to compensate for the surface deformations, or the extent of the mid-IR  surface would have to be reduced. Another option would be to make the mid-IR  surface supported independently of the submm panelized structure.
To minimize structural modifications, the optical design would not be changed, and thus the same subreflector would be used at both mid-IR and submm wavelengths. This implies that if the innermost 3 m of the primary will be used in the mid-IR then the original Cassegrain focal ratio of the ALMA antenna would become F=26.35 at these wavelengths. Only the innermost 21 cm of the secondary would thus be used, though the blocking effciency (86$\%$) would still be determined by the total aperture of the subreflector (80 cm). The Cassegrain plate-scale  (2.6 arcsec/mm, with a 26.35 focal ratio at mid-IR wavelengths) will be converted through the use of relay optics;  for example, if the selected detector focal ratio were F$_{det}$=3.83 then the plate-scale would be 0.54 arcsec/mm  (InSb 256 x 256 detector) and 1.35 arcsec/mm (Si:As $128\times128$ detector) at 3 and 20 $\mu$m, respectively.

Given the novel structure of the primary surface (a central surface with RMS $< 0.5$ micron, surrounded by a panel-ized surface with RMS $< 12$ micron) it is important to analyze the quality of the resulting PSF at mid-IR wavelengths, if the whole antenna surface were used in the mid-IR  to maximize the collecting power (rather than just the inner surface). This has been done assuming that the detectors are incoherent and thus simulating the resulting PSF by summing the PSF  produced by a (assumed perfect) central 3m diameter surface, and that  produced by a surrounding "ring" extending up to a radius of 6m, with a surface RMS of about 12 micron. It can be seen in Figure 12 that the contribution to the PSF by the surface annulus between radii 1.5 and 6m has a central narrow peak surrounded by an irregular error-beam. However, the modifications required to the antenna structure by the mid-IR  option are of 2nd order compared to the ÒsimpleÓ winterization of the antenna, and would have to be analyzed carefully and compared to other options for a larger facility (e.g., a FIR interferometer).

In order to start analyzing all technical issues involved, working meetings have been held with EIE and Thales Alenia Space to investigate possible collaborations with the industry and to discuss some of the modifications required for the winterization of the ALMA antenna. 

\begin{figure}
\centering
\includegraphics[scale=0.45]{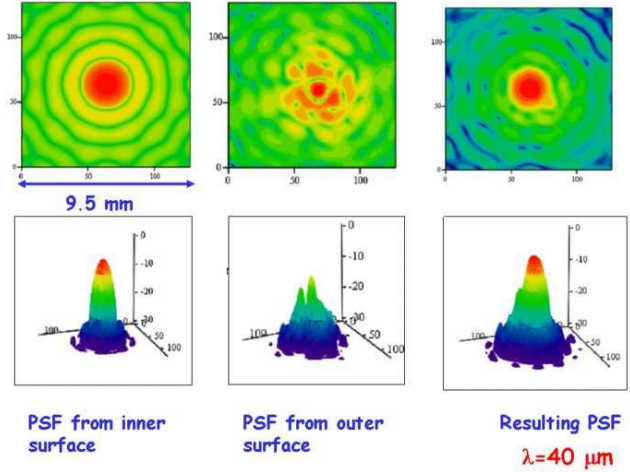}
\caption{Simulated PSFs at 40 micron. Left panels: PSF generated by a perfect 3-m diameter surface, normalized to the peak intensity (top) and in dB below the peak of the total PSF (bottom).  Center panels: PSF generated by the surface annulus between radii 1.5 and 6m with a 12 micron RMS. Right panels: Total PSF. In the top panels, the first contour level and the spacing between adjacent levels is -10dB. The size of the plotted area is $9.5\times9.5$ mm2.}
\end{figure}

\section{Conclusions and future work}

Based on our ARENA prospective activity in defining and identifying unique science cases for FIR/submm astronomy, large ($\sim25$ m) telescopes are essential to make a big step forward toward understanding the cold Universe. This work will now proceed through an ARENA Working Group specifically dedicated to submm/FIR astronomy. Science cases being listed, our future studies will address major questions:
\begin{itemize}
\item Is Dome C really the best site on Earth for the installation of a large submm-wave telescope ? The current answer is  that Dome C is not the best site in Antarctica, Dome A being probably a better site. However, if Dome C is better than any Chajnantor site and the South Pole in terms of overall transmission, the growing infrastructure of Concordia would make possible the deployment of a large telescope facility.  This is not the case at Dome A where no infrastructure is available. How does the polar environment impact hardware? A site testing campaign in wintertime has specifically been designed for FIR/submm astronomy. Three prerequisites are now tested at Dome C before launching (or not) any advanced concept study: (1) whether the submm/THz atmospheric windows open from 200 micron during a large and stable fraction of time; (2) the knowledge of the thermal gradient and (3) icing formation and their impact on a telescope mirror and hardware. In parallel atmospheric modelling and meteorological data analysis will be carried out.
\item Which observatory, which telescope, and how to operate it ? The ARENA WG will study the possibility and the feasibility of installing a large ($>10$ m) single-dish antenna at Dome C for a dedicated project or as an international facility, including: (1) a pre-study on the polar winter impact on telescope infrastructures; (2) a plan for the building and installation of the telescope that takes into account the logistics issues; (3) the development of a power supply respectful of the Antarctic ecology; (4) a cost estimate for the telescope and its operation. All these will be achieved in partnership with industries (e.g. Thales Alenia Space and EIE), CEA laboratories within INES, IPEV (the French polar institute), PNRA (the Italian polar institute) and academic partners within ARENA. 
\end{itemize}

Final conclusions will be presented at the Third ARENA conference in 2009.

\begin{figure}
\centering
\includegraphics[scale=0.3]{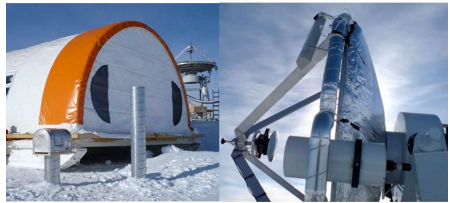}
\caption{Left: A submm tipper experiment to measure the transmission at 200 $\mu$m. Right: GIVRE on Cochise: The process of ice formation and the way of protecting the instruments are described. A 2.6-m telescope (COCHISE) will be defrosted using several methods. GIVRE will also measure other parameters that are common prerequisites for the design of unmanned telescopes at altitudes up to 45 m including the ongoing measurements of wind, humidity, and temperature.}
\end{figure}


\end{document}